\documentclass[prl,twocolumn,amsmath,amssymb,showpacs]{revtex4}
\usepackage{graphicx}
\usepackage{dcolumn}
\usepackage{bm}
\usepackage[usenames]{color}

\usepackage[all]{xy}
\usepackage{amsmath}
\usepackage{natbib}
\def \bea{\begin{eqnarray}}
\def \eea{\end{eqnarray}}

\begin{document}

\title{Effective Temperature of Red Blood Cell Membrane Fluctuations}
\author{Eyal Ben-Isaac$^1$, YongKeun Park$^2$, Gabriel Popescu$^3$, Frank L.H. Brown$^4$, Nir S. Gov\footnote{nir.gov@weizmann.ac.il}$^1$ and Yair Shokef\footnote{yair.shokef@weizmann.ac.il}$^5$}
\affiliation{
$^1$Department of Chemical Physics, The Weizmann Institute of Science, Rehovot 76100, Israel\\
$^2$ Department of Physics, Korea Advances Institute of Science and Technology, Daejeon 305-701, Republic of Korea\\
$^3$ Quantitative Light Imaging Laboratory, Department of Electrical and Computer Engineering, Beckman Institute for Advanced Science \& Technology, University of Illinois at Urbana-Champaign, Urbana, IL 61801, USA\\
$^4$ Department of Chemistry and Biochemistry and Department of Physics, University of California Santa Barbara, CA 93106, USA\\
$^5$Department of Materials and Interfaces, The Weizmann Insitute of
Science, Rehovot 76100, Israel}

\begin{abstract}

Biologically driven non-equilibrium fluctuations are often
characterized by their non-Gaussianity or by an ``effective
temperature'', which is frequency dependent and higher than the
ambient temperature. We address these two measures theoretically by
examining a randomly kicked ``particle'', with a variable number of
kicking ``motors'', and show how these two indicators of
non-equilibrium behavior can contradict. Our results are compared
with new experiments on shape fluctuations of red-blood cell
membranes, and demonstrate how the physical nature of the motors in
this system can be revealed using these global measures of
non-equilibrium.

\end{abstract}

\pacs{87.10.Mn,87.16.D-,87.16.Ln,05.40.-a}

\maketitle


Experimental and theoretical studies of biological systems
confront the issue of active elements which give rise to
fluctuations that distinguish living matter from inanimate
soft-matter systems. Examples range from molecular motors in the
cytoskeleton~\cite{mckintosh} and active membrane
pumps~\cite{patricia} to larger scale objects such as swimming
bacteria~\cite{bacteria}. As with other non-equilibrium systems such
as driven granular matter, it is unclear how to define useful
measures for non-equilibrium ``activity''. Spontaneous fluctuations
may be compared with the response of the system to small external
perturbations, to define an effective temperature $T_{\rm eff}$
using the fluctuation-dissipation (FD) formalism~\cite{Teff}. In most
cases $T_{\rm eff}$ is frequency dependent (unlike the thermal
case), and is larger than the ambient temperature. These features
quantify the non-thermal activity in the system.
Another parameter that is useful for
characterizing deviations from equilibrium is the non-Gaussianity
$\kappa$ of the distribution
function~\cite{gran_kurt,brangwynne,popescu}.

In biological systems the nature of the
microscopic active elements is difficult to study directly.
We demonstrate how global statistical measures of
activity can be used to extract qualitative and quantitative
properties of the underlying molecular motors. We are
motivated by living cells, in which the activity
is induced by multiple motors throughout the system that are
directly coupled to local degrees of freedom. It is important to
explore the effects of the number of motors and their level of
activity on the non-equilibrium nature of the
fluctuations~\cite{pisman_leticia_levine}.

One well-studied active system is the membrane of the
red-blood cell (RBC). However, the nature of the molecular motors in
this system is still far from being well understood. We question how
$T_{\rm eff}$ and $\kappa$
correlate (or not) with each other, and which properties of the
non-equilibrium system do they probe. We focus on RBC and present
new experimental measurements, but obtain results on the
non-equilibrium statistical mechanics of active systems in general.
We introduce a simple model of a randomly kicked ``particle'', with
a variable number of kicking ``motors'' (force producing elements).
Our generalized particle and motors may represent different objects
in different systems, and we will be more specific when comparing to
RBC experiments. We compute both $T_{\rm eff}$ and $\kappa$, and
identify situations in which these two
non-equilibrium indicators contradict.


\emph{Model.} We consider the following overdamped Langevin
equation for the velocity $v=\dot{x}$,
\bea
\dot{v} = - \lambda v + f_T + f_A + f_R . \label{eq:langevin}
\eea
$\lambda$ is the damping coefficient. The thermal force $f_T(t)$ is
an uncorrelated Gaussian white noise: $\langle f_T(t)f_T(t')
\rangle=2\lambda T_B\delta(t-t')$, with $T_B$ the ambient
temperature, and Boltzmann's constant set to $k_B=1$. For the active
force $f_A(t)$ we assume that each of the $N_m$ motors produces
pulses of force $\pm f_0$, of duration $\Delta \tau$.
We assume symmetry with respect to the force direction,
which is motivated by experimental observations of nearly
symmetric active fluctuations of cells~\cite{popescuSymmetric}.
While the pulses turn on
randomly as a Poisson process with an average waiting time $\tau$,
unless otherwise stated, we take a constant pulse length
$\Delta\tau$. The power stroke of molecular motors is a realization
for such a relatively well-defined impulse length~\cite{motors}. We
will also consider stochastic pulse lengths with an arbitrary
distribution $P(\Delta\tau)$, and show that if $P(\Delta\tau)$ is
Poissonian, the force correlations reduce to the shot-noise form
studied in~\cite{nir2004}.

%
To measure the linear response $\chi_{xx}$ of the particle position
$x$, we apply a small force $f_R=F_0 e^{i \omega t}$, and find that
the position is perturbed as $\langle \delta x(t) \rangle =
\chi_{xx}(\omega) F_0 e^{i \omega t}$, with $\chi_{xx}(\omega) =
(\omega(i\lambda - \omega))^{-1}$, $\delta x$ being the small change
in position, irrespective of the active force~\cite{SI}.


Due to the linearity of Eq.~(\ref{eq:langevin}), and since $f_T$ and
$f_A$ are uncorrelated, the velocity autocorrelation is~\cite{SI}
\bea S_{vv}(\omega)  = \frac{2\lambda T_B}{\lambda^2+\omega^2}
+\frac{2N_mv_0^2\lambda^2\left[1-\cos(\omega\Delta\tau)\right]
}{(\tau+\Delta\tau)\omega^2(\lambda^2+\omega^2)}. \label{eq:Svv}
\eea
where $v_0=f_0/\lambda$ is the asymptotic velocity that the particle
approaches due to the activity of a single motor.


\emph{Effective Temperature.} In equilibrium, the
FD theorem connects the imaginary part of
$\chi_{xx}$ to the position autocorrelation,
$S_{xx}(\omega) = - \omega^{-2}S_{vv}(\omega)$,
by: ${\rm Im}\left[\chi_{xx}(\omega)\right] = \frac{\omega}{2T}
S_{xx}(\omega)$. For non-equilibrium steady states we define a
frequency-dependent effective temperature
\bea T_{\rm eff}(\omega) \equiv \frac{\omega S_{xx}(\omega)}{2 {\rm
Im}\left[\chi_{xx}(\omega)\right]} = T_B +
\frac{N_mv_0^2\lambda\left[1-\cos(\omega\Delta\tau)
\right]}{(\tau+\Delta\tau)\omega^2} . \label{Teff} \eea
Note that the time between pulses enters only through the density of
pulses per unit time. Hence this result does not depend on the
distribution of waiting times between pulses, but only on its
average, $\tau$. Moreover, our results may be extended to a stochastic pulse
length~\cite{SI}. In particular, for Poissonian $\Delta\tau$
we obtain shot-noise force correlations $\langle
f_A(t)f_A(0) \rangle=\langle\Delta
\tau\rangle^{-1}\exp{(-t/\langle\Delta \tau\rangle)}$ and $T_{\rm
eff}= T_B + N_mv_0^2\lambda (\tau+\langle\Delta\tau\rangle)^{-1}
\left(\omega^2+\langle\Delta\tau\rangle^{-2}\right)^{-1}/2$.
Note that alternative definitions of the effective temperature have
appeared in the context of granular gases \cite{shokef}. Unlike
those systems, here we do not identify a non-equilibrium situation
where $T_{\rm eff}(\omega)={\rm const}$.
In the high-frequency limit the active contribution vanishes, so
that $T_{\rm eff}\rightarrow T_B$. Around $\omega=1/\Delta \tau$,
$T_{\rm eff}$ rises and as $\omega \rightarrow 0$, it approaches a
constant value in the low frequency limit ($\omega\ll\Delta
\tau^{-1}$): $T_{\rm eff}(0)=T_B+N_m v_0^2\lambda \Delta \tau^2 /
(\tau+\Delta\tau)$.

\begin{figure}
\includegraphics[width=1\columnwidth]{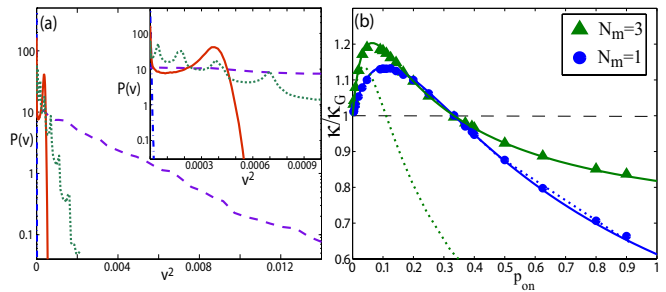}
\caption{(a) Velocity distribution function. Thermal
(dash-dot straight line with slope $T_B^{-1}$ visible in the
inset), $N_m=1$ (solid) and $10$ (dashed).
$\lambda=50$, $\Delta\tau=0.1$,
$\tau=0.15$, $T_B=10^{-4}$, $v_0=0.02$. Note the peak at $v_0$ for
$N_m=1$. Dotted line: $N_m=10$ using $T_{B}=10^{-7}$ and
$\lambda=150$. The kurtosis for the active cases are:
$\kappa/\kappa_{G}=0.85,0.99,0.96$, respectively.
(b) Kurtosis vs motor
activity (varied by changing $\tau$) from numerical simulations
(symbols), compared to the analytic expression ignoring pulse
overlaps (dotted lines), and to the model of shifted Gaussians
(solid lines).}\label{FigPv}
\end{figure}

\emph{Velocity Distribution and Kurtosis.} In Fig.~\ref{FigPv}a we
plot the velocity distribution function $P(v)$ from numerical
simulations. The distribution is highly non-Gaussian for a single
motor. Interestingly, for small $\lambda\Delta\tau$ and in the
presence of multiple motors, $P(v)$ may retain its Gaussian form,
even though the system is very far from equilibrium, as can be seen
in the strong frequency dependence of $T_{\rm eff}(\omega)$, and in
the fact that $\langle v^2 \rangle$ is significantly larger than $T_B$. For our model
we can exactly calculate~\cite{SI}:
$\langle v^2 \rangle=T_B + \left[N_m v_0^2\left(\lambda\Delta\tau+e^{-\lambda\Delta\tau}-1\right)\right]
/\left[\lambda\left(\tau+\Delta\tau\right)\right]$. Let us emphasize that $\langle v^2 \rangle\neq T_{\rm
eff}(0)$, but rather in the limit $\lambda\Delta\tau\rightarrow 0$
we find that $\langle v^2 \rangle=2T_{\rm eff}(0)$.

We measure the non-Gaussianity by the kurtosis, $\kappa \equiv
\langle v^4 \rangle / \langle v^2 \rangle ^2$, which we plot in
Fig.~\ref{FigPv}b as a function of the number of motors and the
activity quantified by the probability of a single motor to be on:
$p_{\rm on}\equiv\Delta \tau/(\tau+\Delta \tau)$. To calculate
$\langle v^4\rangle$ for a single motor, as long as $\lambda \tau
\gg 1$, we ignore overlaps between the contributions of consecutive
pulses~\cite{SI}. A simple model which works rather well at all
number of motors, approximates the distribution as a sum of shifted
thermal Gaussians~\cite{SI}. As long as $\lambda \Delta \tau\gg1$,
the contribution to the velocity distribution due to the rise and
decay before and after each pulse is small, and this model gives a
very good description (see Fig.~\ref{FigPv}b). In the opposite limit
of $\lambda \tau \ll 1$ and $\lambda \Delta\tau \ll 1$ the velocity
distribution approaches a Gaussian. The value of the kurtosis
measures the spread of the distribution; larger values correspond to
a distribution that is wider than a Gaussian, and vice versa.

The most outstanding result is the non-monotonic dependence
of $\kappa$ on $p_{\rm on}$. Compared to a dead system
($p_{\rm on}=0$), as the motor activity is turned on, and when the
active velocity $v_0^2>T_{B}$, the velocity distribution
gets more populated at higher values, hence $\kappa$
increases. In the other limit of $p_{\rm on}\rightarrow1$, $\kappa$
is necessarily smaller than the Gaussian value $\kappa_G=3$, since it is
a contribution of shifted Gaussians. From these
two limits we conclude that $\kappa$ is necessarily a
non-monotonic function of $p_{\rm on}$. In fact, $\kappa$ can
retain its $\kappa_G$ value even for a non-equilibrium
system ($p_{on}>0$). For the distributions shown in
Fig.~\ref{FigPv}a, $\kappa$ is close to $\kappa_G$, except for
the single motor, even when the distribution is visibly non-Gaussian
($N_m=10$). As a function of $N_m$, at small $p_{\rm on}$ the
deviation from $\kappa_G$ increases with $N_m$ (Fig.~\ref{FigPv}b),
since the high velocity tails are more
populated. In the other limit of $p_{\rm on}\rightarrow1$, the
distribution approaches a Gaussian with increasing $N_m$, which is a
manifestation of the central limit theorem.

Comparing with $T_{\rm eff}$ we find that both measures of
non-equilibrium behavior increase with increasing activity in the
$p_{\rm on}\rightarrow0$ limit, while as $p_{\rm on}\rightarrow1$
they contradict. Both $N_m$ and $p_{\rm on}$ are simply multiplied
to give the amplitude of the active contribution to $T_{\rm eff}$,
while $\kappa$ is a more complicated function of these two parameters.


\emph{Experiments.} The activity of the RBC membrane was recently
measured in two different experiments, which found indications for
non-equilibrium fluctuations when the chemical energy source of ATP
is available. Before comparing with our model we note that the
membrane undulations may be described by the following over-damped
analogue of Eq.~(\ref{eq:langevin})~\cite{nirfrankactive},
\begin{equation}
\dot{h_q}=-\lambda_q h_q + \mathcal{O}_q\left[F_T(q,t)+F_A(q,t)\right]
\label{Langevinh}
\end{equation}
where $h_q$ is the amplitude of the membrane deflection at
wavevector $q$, $\lambda_q=\mathcal{O}_q(\bar{\kappa} q^4+\sigma
q^2)$ is the response of the membrane due to the elastic restoring
forces of curvature and tension (with bending modulus $\bar{\kappa}$
and membrane tension $\sigma$), $\mathcal{O}_q=(4\eta q)^{-1}$ is
the Oseen interaction kernel for a flat membrane in free fluid and
$\eta$ the viscosity of the surrounding fluid. The thermal force
satisfies $\langle F_T(q,t)F_T(-q,t') \rangle=2
T_B\mathcal{O}_q^{-1}\delta(t-t')$, and $F_A(q,t)$ is the Fourier
transform of the active force. For the active forces we consider two
cases; a direct force and a curvature-force~\cite{nir2004}, both
with shot-noise correlations.

\begin{figure}
\includegraphics[width=1\columnwidth]{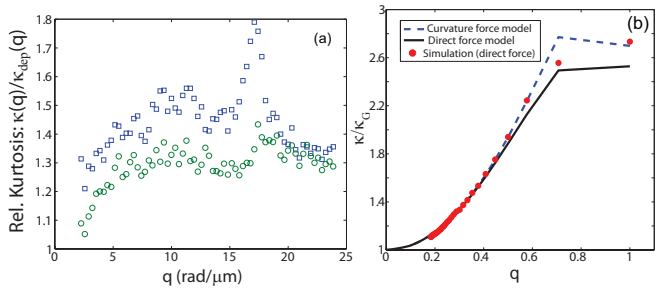}
\caption{(a) Relative kurtosis vs wavevector: Squares (circles) are
for natural (starved for $6$hrs) RBC. $\kappa_{G}(q)$ was
extracted from the data on ATP-depleted cells, starved for $24$hrs.
(b) Calculated dependence of $\kappa$ from our model, mapped to
$q$-space (using $p_{on}=0.07$). The
dimensionless $q$ is determined by varying the
number of motors ($N_m=1,2...30$)~\cite{SI}.}\label{FigPopescu}
\end{figure}

The first experiment \cite{popescu} measured the spatial dependence
of the membrane fluctuations, and extracted the probability
distribution $P(h_q)$, from which the kurtosis was obtained. Here,
$\kappa > \kappa_{G}$ was found for ATP-containing cells. In
Fig.~\ref{FigPopescu}a we present new data showing that $\kappa$
increases with $q$ and with the ATP concentration~\cite{SI}.
Comparing these observations with our model (Fig.~\ref{FigPv}) this
indicates that the RBC has $p_{\rm on}$ close to zero, which means
that $\tau\gg\Delta \tau$.

Next, we compute the $q$-dependence of the kurtosis using our
single-particle model. We map each mode $q$ of
the membrane to a single particle as follows; the number of motors
that act on the membrane area involved in the motion of mode $q$ is
given by: $N_m\propto q^{-2}$ (the number of motors in the
membrane area of wavelength $2\pi/q$, assuming they are uniformly
distributed on the membrane), $F_A\propto q^{0},q^{2}$ (direct and
curvature force respectively~\cite{nir2004}) and $\langle
F_{T}^2\rangle\propto q$~\cite{SI}. Figure~\ref{FigPopescu}b
shows that our calculation predicts that $\kappa\rightarrow\kappa_G$
as $q\rightarrow0$, in agreement with the experiment
(Fig.~\ref{FigPopescu}a). For small $q$, $N_m$ increases,
and we are in the low $p_{\rm on}$ regime. Note that if the
RBC had large $p_{\rm on}$, $\kappa$ would \emph{decrease} with
increasing $q$. The peak in the experimental data may
indicate the wavelength corresponding to a single active unit
(``motor'') in the RBC cytoskeleton \cite{popescu}. Note that using
our single-particle model for the dynamics of an extended
object such as the membrane is only a qualitative approximation.

Thus, the motors are
distributed throughout the RBC membrane, and each motor has a long
recovery time. These findings agree
with the proposed mechanism of membrane fluctuations~\cite{NirSam};
ATP induces the release of membrane-anchored filaments, the
release event ($\Delta \tau$) is fast compared to the time it takes
the released polymer to find its anchor on the membrane and
re-attach ($\tau$).

\begin{figure}
\includegraphics[width=1\columnwidth , bb=0 0 402 207]{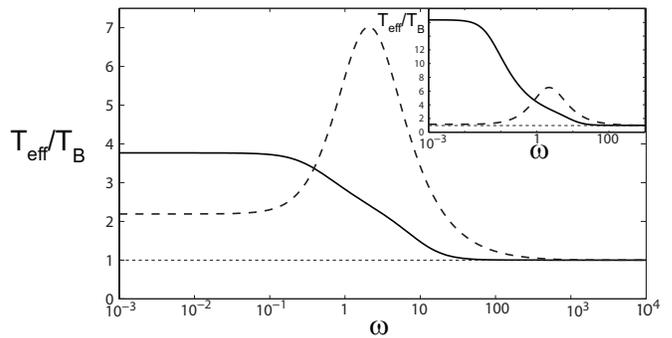}
\caption{Effective temperature for a free membrane of lateral size
$L=8\mu$m, driven by the direct force (solid line) and curvature
force (dashed line). Inset: $L=50\mu$m.}\label{FigMembrane}
\end{figure}

Another experiment~\cite{timo} measured the frequency dependence
of the height fluctuations at a single point on the RBC membrane,
and found a 3-7 fold increase in low frequency ($f<10$Hz)
fluctuations compared to cells depleted of ATP.
The way to decouple the ATP-induced changes to the elastic
moduli~\cite{NirSam} from the increase in $T_{\rm eff}$ is
to measure the response in addition to the fluctuations, and this
awaits future experiments. In Fig.~\ref{FigMembrane} we plot the calculated effective
temperature of the system, as defined by Eq.~(\ref{Teff})~\cite{SI}.
$T_{\rm eff}$ approaches $T_B$ for large frequencies
$\omega\gg\tau^{-1}$, and increases for frequencies
$\omega\leq\Delta \tau^{-1}$. There is even a peak in $T_{\rm eff}$
for the curvature-force. The values of $T_{\rm eff}$ reach up
to $10T_B$, and depend on the lateral size of the membrane $L$; in the
$\omega\rightarrow 0$ limit we find that $T_{\rm
eff,direct}-T_B\propto q_{\rm min}^{-1}$ while $T_{\rm
eff,curv}-T_{B}\propto q_{\rm min}$, where $q_{\rm min}=2\pi/L$
(Fig.~\ref{FigMembrane} inset).
By comparing the calculated and observed \cite{timo} frequency
dependence of $T_{\rm eff}$ and the power spectral density
\cite{SI}, we can estimate the properties of the active ``motors''
in this system: $\Delta \tau\simeq100$msec,
$p_{on}f_0^2\simeq(\bar{\kappa}/r)^2$, where $r\simeq100$nm,
$\bar{\kappa}\simeq90$k$_{B}$T, and the motor density $n=1/r^2$.
These parameters agree with the physical interpretation of the
active force as arising from ``pinching'' of the membrane by a
cytoskeleton network of spectrin filaments \cite{NirSam}.

We demonstrate in Fig.~\ref{FigMembrane} that a future
measurement of $T_{\rm eff}$ can be used to distinguish between
different models of active forces in the membrane, and can therefore
add important information about the physical nature of the motor
producing these forces. In particular, a non-monotonic
behavior will favor the curvature mechanism, while a simpler
step-like behavior will support the direct force.
For the curvature force, the effective temperature
decreases with decreasing $\omega$, and even approaches the
equilibrium value at $\omega\rightarrow0$ for a large membrane
domain ($L\rightarrow\infty$). Another indication for an increase of
effective temperature with frequency was found for a driven granular
system~\cite{granular}. For the RBC membrane this
behavior is driven by the fact that the curvature force couples to
the fluctuation modes of the membrane through a $q^2$
term~\cite{nir2004}, which represents in $q$-space the force due to a
localized induced curvature. This force therefore diminishes in its
relative amplitude as the wavelength increases, leading to the
vanishing of the active component in the
$\omega\rightarrow0$ limit.
We expect this feature to appear in many
spatially extended systems where the driving force
decreases with increasing wavelength.


\emph{Conclusion.} We presented a simple model for active
systems, for which we can derive two measures to characterize its
non-equilibrium nature. These two measures do not always agree.
In biological systems the activity is often
driven by multiple molecular motors that couple to
internal degrees of freedom. In such systems we explored the
characteristics of the non-equilibrium fluctuations in the presence
of multiple motors. We find that in the limit of many
motors the kurtosis can return to the value of a Gaussian
distribution, while the effective temperature may still exhibit
strong frequency dependence. We showed that Gaussian distributions
may arise for active systems even for a small number of motors,
while large deviations from Gaussian distributions can be maintained
even for large number of motors. Note that a non-Gaussian
distribution is not a proof of non-equilibrium, as it could also
arise due to nonlinearities in a mechanical system, such as position
dependent damping. The effective temperature and the kurtosis that
we calculated are explicitly dependent on the number of motors
($N_m$) and their intrinsic properties ($f_0,\tau,\Delta\tau$). Our
present analysis gives a detailed and general treatment, for any
type of pulse-length distribution.

Finally, we compared the results of our model with recent
observations of ATP-driven activity in RBC, and demonstrated how
they can give insight to the underlying active mechanism. In
particular, we showed how fundamental physical properties of the
elusive molecular motor of the RBC membrane may be unraveled by
comparing these observables with our theoretical model. Future
experiments could use the calculated properties to better
characterize the nature of the active forces in various cellular
membranes. We expect our results to be useful for the analysis of
other active systems, both biological~\cite{brangwynne} and
non-biological~\cite{durian,Bocquet}.
From our model we reach the
following more general conclusion: when a spatially extended system
is driven by external forces, the effective temperature defined
through the FD relation can be a
\emph{non-monotonic} function of the frequency. If the coupling of
the external active forces is stronger for smaller wavelengths, then
the effective temperature may develop a non-monotonous dependence on
frequency.

{\bf Acknowledgments:} Y.S. thanks ISF grant 54/08 for support.
N.S.G. thanks the Alvin and Gertrude Levine Career Development
Chair, BSF grant No. 2006285 and the Harold Perlman Family for their
support. GP is partly supported by National Science Foundation
(CAREER 08-46660) and the National Cancer Institute (R21
CA147967-01).

\end{document}